\documentclass[journal,twoside,web]{ieeecolor}
\usepackage{tmi}
\usepackage{cite}
\usepackage{amsmath,amssymb,amsfonts}
\usepackage{algorithmic}
\usepackage{graphicx}
\usepackage{textcomp}

\usepackage{hyperref}       
\usepackage{url}            
\usepackage{booktabs}       
\usepackage{multirow}
\usepackage{mathtools}
\usepackage{verbatim}

\DeclareMathOperator*{\argmin}{argmin}

\begin{document}
\title{Pyramid Convolutional RNN for MRI Image Reconstruction}

\author{Eric Z. Chen,
        Puyang Wang, 
        Xiao Chen,
        Terrence Chen, 
        and Shanhui Sun
\thanks{E. Chen, X. Chen, T. Chen and S. Sun are with United Imaging Intelligence, Cambridge, MA 02140. (e-mail: shanhui.sun@uii-ai.com)}
\thanks{P. Wang is with the Department
of Electrical and Computer Engineering, Johns Hopkins University, Baltimore, MD 21218.}
\thanks{Contribution from P. Wang was carried out during his internship at United Imaging Intelligence, Cambridge, MA.}
}

\maketitle

\begin{abstract}
Fast and accurate MRI image reconstruction from undersampled data is crucial in clinical practice. Deep learning based reconstruction methods have shown promising advances in recent years. However, recovering fine details from undersampled data is still challenging. In this paper, we introduce a novel deep learning based method, Pyramid Convolutional RNN (PC-RNN), to reconstruct images from multiple scales. Based on the formulation of MRI reconstruction as an inverse problem, we design the PC-RNN model with three convolutional RNN (ConvRNN) modules to iteratively learn the features in multiple scales. Each ConvRNN module reconstructs images at different scales and the reconstructed images are combined by a final CNN module in a pyramid fashion. The multi-scale ConvRNN modules learn a coarse-to-fine image reconstruction. Unlike other common reconstruction methods for parallel imaging, PC-RNN does not employ coil sensitive maps for multi-coil data and directly model the multiple coils as multi-channel inputs. The coil compression technique is applied to standardize data with various coil numbers, leading to more efficient training. We evaluate our model on the fastMRI knee and brain datasets and the results show that the proposed model outperforms other methods and can recover more details. The proposed method is one of the winner solutions in the 2019 fastMRI competition.
\end{abstract}


\IEEEpeerreviewmaketitle

\section{Introduction}

\IEEEPARstart{M}{agnetic} Resonance Imaging (MRI) as a non-invasive approach has many advantages over other imaging modalities. However, the MRI data acquisition is inherently slow. One common approach to accelerate MRI data acquisition is to take fewer measurements, generating an undersampled k-space. To increase the signal-to-noise ratio, parallel imaging with multi-channel receiver coils is routinely used. Reconstructing images from the undersampled data is challenging and has been an active research field. 

Compressed sensing (CS) lays theoretical foundations for fast MRI image reconstruction, which includes two important elements \cite{lustig2008compressed}: the sparsity regularization to constraint the solution space \cite{ma2008efficient,ye2019compressed} and the optimization algorithms to find the optimal solutions \cite{chambolle2011first,wang2008new,ramani2010parallel,ye2019compressed,fessler2019optimization}. Both the sparsity regularization and the optimization algorithms are often manually designed.  As data-driven approaches,  deep learning (DL) based methods have been proposed for MRI reconstruction in recent years and show promising improvements over CS \cite{sandino2020compressed,liang2020deep}. 

However, it is still a challenge for current DL methods to recover high frequency signals (i.e., fine details) from undersampled data, especially with a high acceleration rate \cite{hyun2018deep,chen2020real,seitzer2018adversarial,liang2020laplacian,deora2020structure}. The DL based MRI reconstruction methods tend to generate over-smoothed (i.e., blurry) images,  which is due to the absence of high frequency information \cite{chen2020real,qin2020deep,sriram2020grappanet,mardani2017deep,liang2020laplacian,zhang2020compressed,lebel2020performance,dar2020prior}. The high frequency information affects not only the sharpness of the reconstructed image, but also the small pathology in the reconstructed image \cite{knoll2020advancing,sriram2020grappanet}. Therefore, recovering the high frequency information in the reconstructed image is crucial for clinical diagnosis. To improve the sharpness, different methods have been proposed such as GAN \cite{dar2020prior,quan2018compressed,yang2017dagan,deora2020structure}, cascade models \cite{schlemper2017deep,eo2018kiki}, cross-domain learning \cite{eo2018kiki}, residual learning \cite{lee2018deep} and high frequency learning \cite{zhang2020compressed,he2019learning}. However, those methods either do not explicitly model the high frequency information or model the low and high frequency information in a parallel fashion. Furthermore, the GAN based methods may introduce undesired artifacts \cite{qin2020deep,tan2021systematic,abdal2019image2stylegan,demir2018patch,zhu2020gan,lucas2019efficient,chen2021hierarchical}, which is unfavorable in the clinical setting. 

In order to improve the image reconstruction quality, especially the fine details, we propose a pyramid convolutional RNN (PC-RNN) model to learn multi-scale features based on two motivations. First, the challenge to recover high frequency signals in MRI reconstruction can be explained by the spectral bias of DL models \cite{rahaman2019spectral,xu2019frequency,ronen2019convergence,cao2019towards}. The neural networks, in general, tend to learn the low frequency pattern first, which leads to the bias of generating smooth reconstruction images. Therefore, this motivates us to design a specific network architecture to force the network to explicitly reconstruct the signals from low to high frequency in a pyramid fashion in order to overcome this spectral bias. 

Second, MRI reconstruction often involves iterative optimization algorithms. Based on the idea of learning to optimize \cite{andrychowicz2016learning,li2016learning,chen2021learning,li2017learning,lv2017learning,chen2017learning,ravi2016optimization,dai2017learning,wang2016learning,finn2017meta}, the classical optimization algorithms (e.g., the gradient descent and its variants) can be learned by DL models such as LSTM \cite{andrychowicz2016learning} and deep reinforcement models \cite{li2016learning}.  
During optimization, those conventional algorithms are functions evaluated repeatedly, which can be learned using neural networks by training in a data-driven fashion. The neural networks themselves can be optimized during training by the optimization algorithm such as the gradient descent. Therefore, it is called ``learning to learn by gradient descent by gradient descent" in \cite{andrychowicz2016learning}. Those learned optimizers using DL models outperform the conventional hand-designed methods (e.g., SGD, RMSprop and ADAM) in terms of convergence speed, generalization and final solutions \cite{andrychowicz2016learning, chen2021learning}. 

Therefore, we propose to learn the optimization process in MRI reconstruction by the convolutional RNN (ConvRNN) model. We further extend this idea to model multi-scale information with three ConvRNN modules in a coarse-to-final manner, where each ConvRNN module has a recurrent encoder-decoder architecture with a different downsampling rate.
Based on the coarse output from the previous ConvRNN module, the following ConvRNN module extracts more refined features. The extracted features in coarse to fine scales are aggregated by a final CNN module in a pyramid fashion. The final CNN module provides guidance for the ConvRNN modules to extract complementary features such that the final CNN module generates the best result. This encourages the last ConvRNN module to explicitly learn high frequency features since the first two ConvRNN modules learn features mainly at low and medium scales due to the downsampling in the encoder-decoder. Each ConvRNN and final CNN have data consistency layers to enforce the data consistency at each scale. Hence our method can recover fine details while preserving data consistency.

From the optimization perspective, this can be regarded as sequentially searching for solutions in a coarse-to-fine manner. Our model first searches for a coarse solution in a smaller space. Based on the previous solution, our model then gradually searches the larger space to find the final optimal solution. Hence, with the help of the previous solutions, the optimization does not need to search the whole large space. This multi-scale searching strategy can often lead to faster convergence and better solutions than directly optimizing in the original large space \cite{mjolsness1991multiscale}.  

It is common to utilize coil sensitivity maps (CSM) for multi-coil data from parallel imaging \cite{knoll2020deep}. Several DL-based approaches have been proposed for multi-coil data by pre-computing CSM or learning CSM during the reconstruction. However, it often takes time to calculate CSM and hence is less time efficient, especially for large data. Methods that learn CSM often design separated networks for CSM estimation \cite{sriram2020end,pezzotti2019adaptive} and lead to more GPU memory consumption and training time. On the other hand, the reconstruction quality relies on the accuracy of the estimated CSM. Inaccurate CSM often leads to artifacts in the final coil-combined image \cite{holme2019enlive}.  
Several calibrationless reconstruction methods have also been proposed \cite{liu2021calibrationless,pramanik2020deep,wang2020deepcomplexmri,darestani2021accelerated,pramanik2021joint,schlemper2019data}. However, once those models are determined and trained, they can not be directly applied to the data with a different number of coils. Therefore, to avoid the above limitations, our model takes the multi-coil data as multi-channel inputs without explicitly using CSM. For data with different numbers of coils, we use coil compression \cite{zhang2013coil} to standardize the coil dimension. 

In summary, our main contributions are: 
\begin{itemize}
\item We propose a novel pyramid multi-scale ConvRNN model for MRI reconstruction and explicitly reconstruct MRI images in a coarse-to-fine manner.
\item We propose to model multi-coil MRI data without CSM and apply coil compression to standardize the coil dimension. 
\end{itemize}

\section{Related work}

In this section, we will review DL-based MRI reconstruction methods. Please refer to \cite{ye2019compressed} for CS-based approaches.

Most DL approaches borrow the basic ideas from CS. Some approaches use neural networks to learn the prior (i.e., the sparsity regularization used in CS) from training data, such as variational autoencoder\cite{tezcan2018mr}, GAN \cite{mardani2017deep,yang2017dagan,bhadra2020medical,deora2020structure,seitzer2018adversarial,dar2020prior} and other generative models \cite{luo2020mri,liu2020highly,he2019learning}. Some works proposed to learn a low-dimensional manifold representation of the MRI images, which can then be used as a prior for image reconstruction. For example, \cite{bermudez2018learning} showed that the brain MRI images form a low-dimensional manifold and it can be learned by a DL model. Several CNN based models were developed to learn a smooth low-dimensional manifold for  MRI image reconstruction \cite{biswas2019dynamic,wen2020transform,ke2021deep}. \cite{zhu2018image} proposed a manifold learning approach to learn sparse transforms that lead to the low-dimensional representation of the data. Based on the idea of deep image prior \cite{ulyanov2018deep}, some unsupervised MRI reconstruction methods have been proposed \cite{darestani2021accelerated,yoo2021time,van2018compressed,john2021deep,korkmaz2021unsupervised}. Those methods either do not explicitly learn high frequency prior or learn low/high frequency priors in a parallel fashion.

On the other hand, some DL models are designed based on the optimization algorithms used in CS. For example, ADMM-Net \cite{sun2016deep} substitutes the operations in the  ADMM optimization framework by neural networks and the parameters can be learned during training. Similarly, ISTA-Net \cite{zhang2018ista} replaces all the manually designed parameters in the ISTA algorithm with neural networks. VS-Net \cite{duan2019vs} decomposes the optimization problem into several sub-problems using the variable split method and solves each sub-problem with a neural network. Chen \textit{et al.} \cite{chen2019model} proposed a similar method based on the split Bregman iterative algorithm. Hammernik \textit{et al.} \cite{hammernik2018learning} generalized optimization formulation in CS with a variational network. Since the optimization algorithms are often iterative, many DL models utilize similar architectures to iteratively update the reconstruction results. The most common iterative models are unrolled or cascaded networks \cite{schlemper2017deep,huang2019mri,souza2019hybrid,souza2019dual,eo2018kiki,wang2018dimension,bao2019undersampled,aggarwal2018modl,gilton2019neumann, pezzotti2019adaptive,zhou2020dudornet,zhang2020high,meng2019prior,liu2020rare,diamond2017unrolled,ke2020deep,yazdanpanah2019deep,wang2019lantern,dar2020transfer}, RNN models \cite{qin2018convolutional,oh2021k} and iterative inverse model \cite{putzky2019invert}. Some DL approaches perform MRI reconstruction by learning the direct mapping \cite{lee2018deep,han2019k,akccakaya2019scan,zhu2018image,schlemper2019dautomap}, which can be regarded as a special case of iterative models but with only a single iteration. To enforce the data fidelity during optimization, the data consistency layers are often adopted into the iterative network design \cite{schlemper2017deep,aggarwal2018modl,quan2018compressed,zheng2019cascaded,eo2018kiki,hammernik2021systematic}. Those methods commonly use neural networks to replace certain components in the optimization algorithm while still utilizing the framework of the original optimization algorithm. Instead, our method is inspired by the concept of learning to optimize and the ConvRNN model is employed for MRI reconstruction.  

In the case of parallel imaging, most of the DL based methods utilize the SENSE formulation \cite{pruessmann1999sense} for multi-coil data. One strategy is to first combine the multi-coil undersampled images and then feed the network with coil-combined images for reconstruction \cite{zbontar2018fastmri}. Another strategy is to use pre-computed CSM along with multi-coil data in the model \cite{duan2019vs,hammernik2018learning}. Several methods have been proposed to estimate the CSM from the undersampled data using neural networks \cite{chen2019model,hammernik2019sigma,jun2021joint}. For example, VS-Net pre-computes CSM from undersampled data and includes them in the reconstruction \cite{duan2019vs}. Blind-PMRI-Net jointly learns the image prior and explicitly estimates CSM with an iterative reconstruction algorithm \cite{meng2019prior}.  E2E-VN includes two networks, where one network combines multi-coil data using CSM and refines the combined image, and the other network estimates CSM used in the refinement network \cite{sriram2020end}. 
Chen \textit{et al.} introduced a filter operator on the multi-coil data, which is implemented by convolutional layers, and performs reconstruction by implicitly estimating CSM \cite{chen2019model}.  
$\Sigma$-Net introduces a strategy that ensembles the reconstructions from two networks: the sensitivity network, which combines coils explicitly using pre-computed CSM, and the parallel coil network, which implicitly learns coil combination  \cite{hammernik2019sigma}. Some DL models have also been proposed \cite{sriram2020grappanet,kim2019loraki,akccakaya2019scan} based on GRAPPA \cite{griswold2002generalized}. Other calibrationless methods \cite{schlemper2019data,liu2021calibrationless,pramanik2020deep,wang2020deepcomplexmri,darestani2021accelerated,pramanik2021joint} have also been proposed. For example, Deep-SLR \cite{pramanik2020deep} applies CNN model to learn annihilation relations among undersampled k-space measurements. ComplexMRI \cite{wang2020deepcomplexmri} proposed to stack coil images along the channel dimension and utilize complex valued convolutions in the CNN model for parallel MRI reconstruction. In this work, we will demonstrate that the DL model can directly take the multi-coil data as multi-channel inputs without using CSM and the coil compression technique can be utilized for data with various coil dimensions.

\section{Proposed Method}

In this section, we will first formulate MRI reconstruction as an inverse problem, which can be solved as an optimization problem using the gradient descent algorithm. Inspired by the idea of learning to optimize, we will then show that the gradient descent can be learned by a ConvRNN model. Based on this formulation, we will introduce our model, which features three ConvRNN modules that reconstruct images at multiple scales in a pyramid fashion. 

\subsection{MRI reconstruction as an inverse problem}
The MRI data acquisition can be formulated as follows:
\begin{equation}
y = Ax + \epsilon ~,
\label{eq:1}
\end{equation}
where $x \in \mathbb{C}^M$ is the image to reconstruct, $y \in \mathbb{C}^N$ is the undersampled k-space, and $\epsilon$ is the noise.  
$A$ is the forward operator and often the multiplication of the Fourier transform matrix $\mathcal{F}$, the binary undersampling matrix $D$ and the coil sensitivity matrix $S$. In our work, we do not use CSM and ignore $S$ in the following paper.  
MRI reconstruction can be considered as an inverse problem, in which the inverse process is ill-posed due to the information loss in the forward process as $N < M$.  
A regularization term $R(x)$ needs be added to the objective function:
\begin{equation}
\argmin_x \frac{1}{2}||y-Ax||_2^2 + \lambda R(x).
\label{eq:3}
\end{equation}
In CS, $R(x)$ takes the form of $||\Psi x||_1$, where $\Psi$ is the transformation matrix. This $\ell1$ term forces $x$ to be sparse in the transformed domain. In DL, this regularization function $R(x)$ can be learned from data. Note that the regularization parameter $\lambda$ can be absorbed into $R(x)$ to be learned by the network and we omit it in the following equations.  

\subsection{Convolutional RNN}
Many optimization algorithms can be used to minimize the objective function such as gradient descent, proximal gradient descent and primal-dual optimization algorithms. Without loss of generality, Eqn.~\ref{eq:3} can be minimized using gradient decent in an iterative fashion,  
\begin{equation}
\hat{x}^{(k+1)} =  \hat{x}^{(k)} - \alpha [A^T (A\hat{x}^{(k)}-y) + \nabla R(\hat{x}^{(k)})],
\label{eq:2}
\end{equation}
where $k=0,1,\dots,K$ is the index for iteration and $\alpha$ is the learning rate.

The hand-crafted iterative optimization procedure such as the above gradient descent update rule in Eqn.\ref{eq:2} can be considered as a time series process, which can be learned by LSTM \cite{andrychowicz2016learning}, deep reinforcement learning \cite{li2016learning} or neural ODE \cite{chen2020mri} models. In the domain of learning to optimize, the recurrent model can act as an optimizer (such as gradient descent) that can be used to optimize the target function, where the recurrent model itself can then be optimized by gradient descent in the training stage\cite{hospedales2020meta}. The similarity between the gradient descent update rule and RNN architecture has also been discussed in \cite{nguyen2020momentumrnn}. 
Inspired by the above works, we utilize a ConvRNN model to learn Eqn.\ref{eq:2} as
\begin{equation}
    \hat{x}=G(\Tilde{x}; \theta), 
\label{eq:4}
\end{equation}
where $\Tilde{x}$ is the input image (i.e., undersampled image) 
and $\theta$ are the model parameters. We omit the parameter notation in the following equations for simplicity. In ConvRNN, the matrix multiplications in the vanilla RNN model are replaced with convolution operations for images.

The proposed ConvRNN is designed for static MRI image reconstruction and employs a recurrent encoder-decoder architecture with various downsampling rates for multi-scale learning (see next section),  
which is different from the previously proposed ConvRNN model for dynamic MRI reconstruction \cite{qin2018convolutional}.

\subsection{Multi-scale Pyramid Convolutional RNN}
\begin{figure}[!t]
  \centerline{
  \includegraphics[width=\columnwidth]{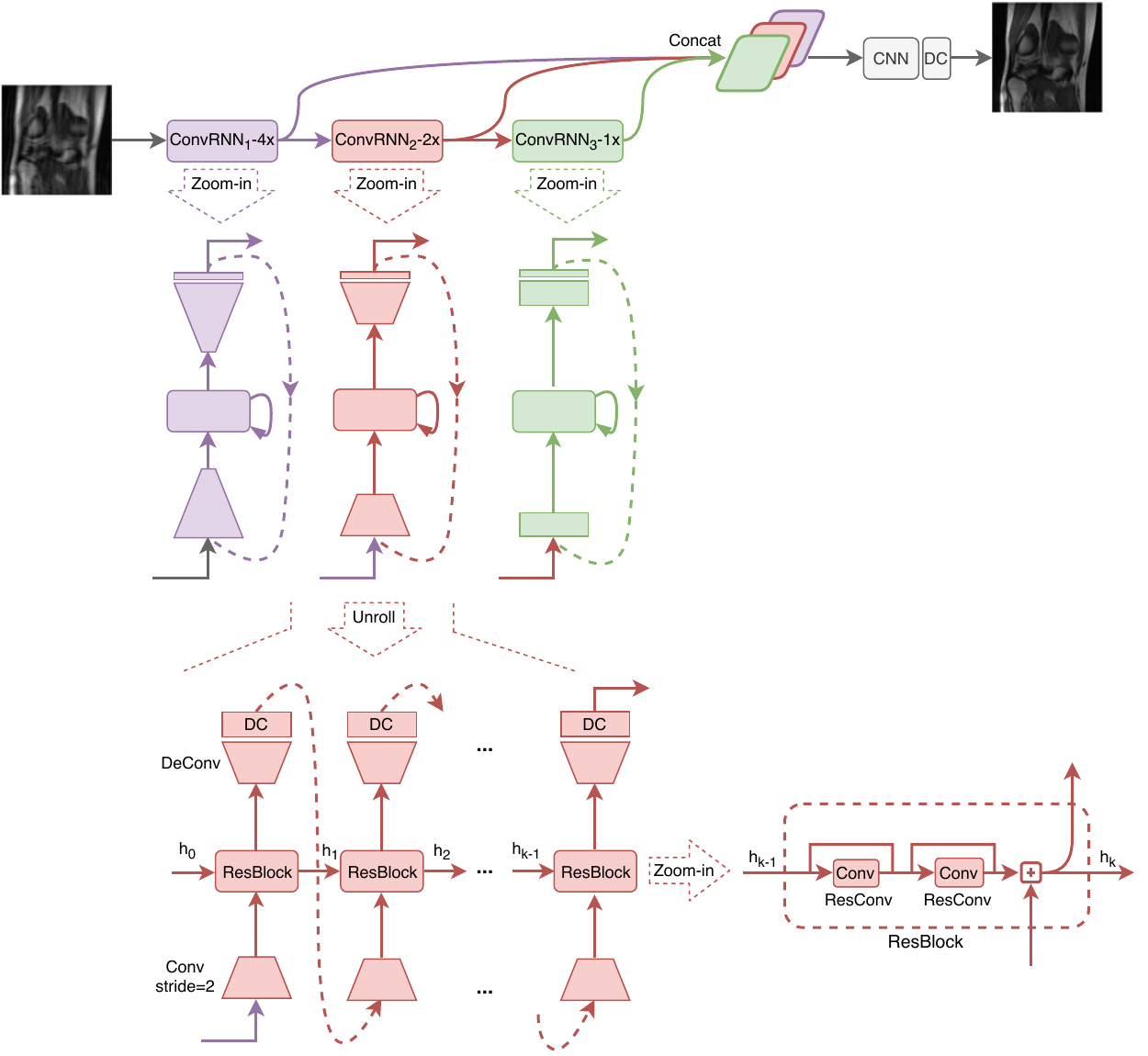}
          }
  \caption{The illustration of the proposed Pyramid Convolutional RNN (PC-RNN) model for MRI reconstruction. The model includes three convolutional RNN modules to iteratively reconstruct images at different scales. The reconstructed images are combined by a final CNN module. }
  \label{fig:pcrnn_model}
\end{figure}

To improve the reconstruction of fine details, we extend the above idea and propose to reconstruct the image at multiple scales. Consider the final reconstructed image $\hat{x}$ as the combination of images $\hat{x}_s$ in different scales.
\begin{align}
\hat{x} &= f(\hat{x}_1, \hat{x}_2, \dots, \hat{x}_S)~,
\label{eq:5}
\end{align}
where $\hat{x}_s$ indicates reconstructed image at scale $s$ ($s=1,\dots,S$ for total $S$ scales) and $f$ is a function to integrate images from different scales. Large $s$ indicates fine scale.

Then we can reconstruct each $\hat{x}_s$ as:
\begin{align}
\hat{x}_s = G_s(\hat{x}_{s-1})~,
\label{eq:6}
\end{align}
where $\hat{x}_0=\Tilde{x}$ is the input image and $g_s$ is the function for reconstruction. Each $G_s$ is specialized in modeling the signal at a certain scale $s$.

The rationale of the above multi-scale learning is that we decompose the original problem of reconstructing MRI images into several sub-problems, in which images at different scales are reconstructed sequentially, from coarse to fine scales, and then combined as the final reconstructed image. In this way, the searching space in each sub-problem learned by $G_s$ (Eqn.~\ref{eq:6}) is smaller and it is easier to find better solutions than the original problem modeled by $G$ (Eqn.\ref{eq:4}). 

We implement the idea from Eqn.~\ref{eq:4}-\ref{eq:6} and propose a novel network Pyramid Convolutional RNN (PC-RNN) as shown in Figure~\ref{fig:pcrnn_model}. It features three ConvRNN modules to model data in different scales, which corresponds to ConvRNN$_1$-4x, ConvRNN$_2$-2x and ConvRNN$_3$-1x in the upper panel of Figure~\ref{fig:pcrnn_model}, 
\begin{align}
    &\hat{x}_1 = G_1(\hat{x}_0,y,D)\\
    &\hat{x}_2 = G_2(\hat{x}_1,y,D)\\
    &\hat{x}_3 = G_3(\hat{x}_2,y,D),
\end{align}
where the undersampled k-space $y$ and the mask $D$ are included for data consistency. 

We apply a final CNN module with four convolutional layers to aggregate the coarse-to-fine features $\hat{x}_1,\hat{x}_2,\hat{x}_3$  in a pyramid fashion and derived the final reconstruction $\hat{x}$:
\begin{equation}
    \hat{x} = f(\hat{x}_1,\hat{x}_2,\hat{x}_3,y,D).
\end{equation}

In PC-RNN, the three ConvRNN modules at different scales are used to process multi-scale information, which is then combined by the final CNN module to generate the final reconstruction. By design, the final CNN module provides guidance for the ConvRNN modules to extract complementary features. The first two ConvRNN modules learn features mainly at low and medium scales due to the downsampling in the encoder-decoder, which encourages the last ConvRNN module to explicitly learn high frequency features. 

To ensure the data fidelity in the original objective function (Eqn.~\ref{eq:3}), the data consistency layers (DC layers in Figure~\ref{fig:pcrnn_model}) are added into each iteration of ConvRNN modules and the CNN module as  
\begin{align}
    \sigma(\hat{x},y,D) = \mathcal{F}^{-1}[Dy+(1-D)\mathcal{F}\hat{x})].
\end{align}

The details of each ConvRNN module $G_s$ are shown in the middle panel of Figure~\ref{fig:pcrnn_model}, which consists of four components: an encoder $g_s^{enc}$, an decoder $g_s^{dec}$, a basic RNN cell (ResBlock) $g_s^{res}$, and a data consistency layer $\sigma$.  The output of $(k+1)^{th}$ iteration of ConvRNN module $G_s$ can be derived as follows:
\begin{align}
    \hat{x}^{(k+1)}_{s} = \sigma(g^{dec}_{s}(g^{res}_{s}(h^{(k)}_{s})+g^{enc}_{s}(\hat{x}^{(k)}_{s}))),
\end{align}
where $h^{(k)}_s = g^{res}_s(h^{(k-1)}_s)+g^{enc}_s(\hat{x}^{(k-1)}_s)$ is the hidden state from previous iteration and $h^{(0)}_s=0$. 

The bottom panel of Figure~\ref{fig:pcrnn_model} shows the details of each component in ConvRNN. To ensure each ConvRNN extracts features at different scales, the spatial sizes of feature maps are downsampled by 4x, 2x, 1x, respectively, using convolutional layers with strides=2 in encoders. The encoder $g_1^{enc}$ in ConvRNN$_1$-4x module includes two convolutional layers with stride=2, which leads to coarse reconstruction at the 4x scale. The ConvRNN$_2$-2x module has one convolutional layer with stride=2 and another one with stride=1 in the encoder, which results in reconstruction at the 2x scale. The last ConvRNN$_3$-1x module has two convolutional layers with stride=1 in the encoder and reconstructs images at the 1x scale. The decoders in each ConvRNN use transposed convolutional (deconvolutional) layers to recover the original image size for the final combination. The ResBlock $g_s^{res}$ includes two residual convolutions (ResConv). 

\subsection{Multi-coil data modeling}
For multi-coil data, the model takes the stack of all coils as multi-channel inputs without using CSM. We combined the real/imaginary channels with coils along the same dimension and fed the data into the network. In the model output, we then separated the real/imaginary channels and coils into two dimensions by reshaping the data. Thus, the network outputs the reconstructions for all coil images. The final reconstruction is obtained by combining multi-coil images using the root sum squared (RSS) method,
\begin{equation}
\hat{x}_{\mathrm{rss}}=\sqrt[]{\sum_{c=1}^{n_{c}}\left|\hat{x}_{c}\right|^{2}}~,
\end{equation}
where $n_c$ is the number of coils. Since the training loss is added onto the combined image instead of the individual coil image, the model learns to weight each coil in an optimal way such that the final combined image matches the ground truth.

\subsection{Coil compression}
Our PC-RNN model does not use CSM and takes inputs with a fixed number of coils. But the coil numbers in the fastMRI brain dataset are different. Therefore, we applied the coil compression technique \cite{zhang2013coil} to standardize the coil dimension in the brain dataset. Since the coil numbers are the same in the knee dataset, we did not employ coil compression on the knee dataset although the methods are still readily applicable. The coil compressed data can be considered as a good approximation to the original true measurements and is used for data consistency.

To perform the coil compression, a $n_{calib} \times n_{calib}$ central region of the k-space of every coil representing low-spatial-frequency component $y_{calib} \in \mathbb{C}^{n_{calib}^2 \times n_c}$ is used as the calibration data. The calibration data is factorized using singular value decomposition (SVD) and the first $n_{vc}$ columns of the right-singular vectors are kept to form a compression matrix $M_c \in \mathbb{C}^{n_c \times n_{vc}}$. 
The acquired $n_c$-coil data is then compressed to $n_{vc}$ virtual coils through $y_{comp} = y M_c$. 

\section{Experiments}

\subsection{Datasets}
We used the knee and brain datasets from fastMRI competition \cite{zbontar2018fastmri}. The knee dataset includes single-coil and multi-coil tasks with 973 volumes (34,742 slices) for training and 199 volumes (7,135) for validation. The brain dataset only includes the multi-coil task with 4,469 volumes (70,748 slices) for training and 1,378 volumes (21,842 slices) for validation. The fully sampled k-space data are available in both training validation data.
Only undersampled k-space data (i.e., no fully sampled data) are provided in the test dataset. For the knee dataset, we used the ground truth images provided by fastMRI. However, for some data in the brain dataset, the image size of ground truth is smaller than the size of the corresponding k-space and the small size is not a center crop of large size. Instead, we regenerated the ground truth images using the inverse fast Fourier transform (iFFT) of the k-space. 

To optimize the model hyper-parameters, we randomly sampled a small dataset, including 100 volumes for training and 50 volumes for validation, from the training data of fastMRI multi-coil knee and brain datasets, respectively. 

\subsection{Training and evaluation}
We combined the Normalised Mean Square Error (NMSE) loss \cite{zbontar2018fastmri} and the Structural Similarity Index (SSIM) loss \cite{zhao2015loss} on the coil-combined images for training, 
\begin{align}
    &\mathcal{L}(\hat{x},x) = \mathcal{L}_{\text{NMSE}} + \beta \mathcal{L}_{\text{SSIM}},
\end{align}
where we empirically set $\beta=0.5$ to balance the two loss functions. 

For both single-coil and multi-coil tasks, we trained models with the same network architecture except for the input and output dimensions and the number of feature maps of each module.
The number of iterations for all ConvRNN is set to 5. We followed the data preprocessing procedure in fastMRI \cite{zbontar2018fastmri} except for the data normalization. All training images were center cropped to $320\times320$ if possible. For images with sizes less than $320\times320$, we center cropped the images and the ground truth to the minimum size of the two data. The undersampled image and undersampled k-space were normalized by dividing by the mean of the magnitude of the undersampled image. The reconstruction results were then transformed back into their original scale (by multiplying the previous mean value) for evaluation. The real and imaginary values of input data are split into two channels. We used the lookahead version of Adam optimizer \cite{zhang2019lookahead}. The learning rate was set to $10^{-5}$ for the training warmup and increased to $10^{-4}$, which was then reduced by a factor of 2 every 10 epochs. The network was trained for 60 epochs with a batch size of 8. We applied the random sampling to the knee dataset and the equispaced sampling to the brain dataset according to \cite{zbontar2018fastmri}. 

We calculated PSNR and SSIM on the validation data for comparison. For a fair comparison, we adopted a similar evaluation approach as in \cite{hammernik2021systematic,muckley2021results}. That is, when evaluating the results (i.e., calculating the PSNR and SSIM), we generated masks from CSM by simply thresholding the RSS of CSM and applied these masks to reconstructed images and ground truth. In this way, all models were evaluated using the same image regions. The Wilcoxon signed-rank test was used to calculate p values to indicate the statistical significance of the difference. 

We compared our method with CS\footnote{https://github.com/facebookresearch/fastMRI} \cite{lustig2008compressed}, U-Net\footnotemark[1] \cite{ronneberger2015u}, D5C5\footnote{https://github.com/js3611/Deep-MRI-Reconstruction}\cite{schlemper2017deep}, KIKI-Net\footnote{https://github.com/zaccharieramzi/fastmri-reproducible-benchmark}\cite{eo2018kiki}, VN\footnote{https://github.com/rixez/pytorch-mri-variationalnetwork}\cite{hammernik2018learning}, VS-Net\footnote{https://github.com/j-duan/VS-Net}\cite{duan2019vs} and ComplexMRI\footnote{https://github.com/CedricChing/DeepMRI}\cite{wang2020deepcomplexmri}. D5C5, KIKI-Net and ComplexMRI are three state-of-the-art methods utilizing the iterative network architecture to improve the reconstruction. KIKI-Net uses a cross-domain strategy that can improve the sharpness of the reconstructed images\cite{eo2018kiki}. ComplexMRI utilized the complex convolution to process the complex values in the MRI data. For multi-coil data, we modified input channels of U-Net, D5C5, and KIKI-Net to take the multi-coil data as the multi-channel inputs. VN and VS-Net are two recent DL models for parallel imaging and show superior performance than previous methods\cite{hammernik2018learning,duan2019vs}. For VN and VS-Net, the models used CSM, which was pre-computed using BART \cite{uecker2015berkeley}. We used the same parameters in \cite{zbontar2018fastmri} for CSM computation (bart ecalib -m1 -r26) and CS reconstruction (bart pics -d4 -i 200 -R T:7:0:0.05). We controlled the model complexity by adjusting the number of convolution channels in each model such that all models have a similar number of parameters. After the adjustment, the number of parameters (in millions) for each model is U-Net (1.1M), KIKI-Net (1.2M), D5C5 (1.3M), VS-Net (1.6M), VN (1.3M) and ComplexMRI (1.6M).  For a fair comparison, we downsized the original PC-RNN-B (23M) used in the fastMRI competition to a smaller version PC-RNN-S (1.6M), where B and S indicate big and small versions (hereinafter, if not specified, PC-RNN refers to PC-RNN-B). We trained separated models for 4X and 6X acceleration in each task. All models were trained and evaluated using the same procedure.

\section{Results}

\subsection{fastMRI knee dataset}

\begin{table}
  \caption{Evaluation results on the validation dataset of fastMRI knee single-coil task.}
  
  \centering
  \begin{tabular}{ccccccc}
    \toprule
     &&&\multicolumn{2}{c}{PSNR}  & \multicolumn{2}{c}{SSIM}\\
    \cmidrule(r){4-5} \cmidrule(r){6-7}
    Task & Sequence  & Method &  4X     & 6X  & 4X     & 6X \\
    \midrule
     &   & CS &  31.4   & 29.4 & 0.645 & 0.601  \\
     &  & U-Net &   33.7  & 32.4 & 0.800 &    0.765   \\
     & PD  & KIKI-Net &   34.2  & 32.5 & 0.814 & 0.775  \\
     &   & D5C5 &  34.3   & 32.6 & 0.817  & 0.778  \\
     &   & PC-RNN-S & 34.9   & 33.6 & 0.828  & 0.796  \\
     &   & PC-RNN-B &  \textbf{35.2}  & \textbf{34.2}  & \textbf{0.835} & \textbf{0.807}    \\
     \cmidrule(r){2-7}
     &   & CS &  27.7   & 27.0 & 0.493 & 0.441  \\
    Knee &  & U-Net & 29.6    & 28.9 & 0.620 &   0.570    \\
    Single-coil &  PDFS & KIKI-Net &  29.9   & 29.1 & 0.631 & 0.580  \\
     &   & D5C5 &  30.0   & 29.2 & 0.635  & 0.584 \\
     &   & PC-RNN-S &  30.2   & 29.5 & 0.645  & 0.594 \\
      &   & PC-RNN-B &  \textbf{30.3}  & \textbf{29.7}  & \textbf{0.651} & \textbf{0.604}    \\
     \cmidrule(r){2-7}
     &   & CS &    29.5 & 28.2 & 0.570 & 0.521  \\
     &  & U-Net & 31.7   & 30.6 & 0.710 & 0.668       \\
     & All  & KIKI-Net & 32.1  & 30.8 & 0.723  &  0.678 \\
     &   & D5C5 & 32.2  & 30.9 & 0.727  & 0.681  \\
     &   & PC-RNN-S & 32.6  & 31.5 & 0.737  & 0.696  \\
     &   & PC-RNN-B &  \textbf{32.8}  & \textbf{31.9}  & \textbf{0.743} & \textbf{0.706}    \\
    \bottomrule
  \end{tabular}
  \label{table_knee_validation_sc}
\end{table}

\begin{table}
  \caption{Evaluation results on the validation dataset of fastMRI knee multi-coil task.}
  
  \centering
  \begin{tabular}{ccccccc}
    \toprule
     &&&\multicolumn{2}{c}{PSNR}  & \multicolumn{2}{c}{SSIM}\\
    \cmidrule(r){4-5} \cmidrule(r){6-7}
    Task & Sequence  & Method &  4X     & 6X  & 4X     & 6X \\
    \midrule
    &   & CS &   32.1  & 28.8 &  0.885 & 0.825  \\
    &  & U-Net &  35.5   & 33.3 & 0.911 &  0.880     \\
    &  & VN &  36.4   & 33.0 & 0.921 &  0.874     \\
    &  & VS-Net &  37.1   & 34.3 & 0.930 &  0.896     \\
     &  PD & KIKI-Net &   38.3 & 35.8 &  0.939 & 0.913  \\
     &   & D5C5 &   38.8 & 35.9 &  0.943 & 0.913  \\
     &   & ComplexMRI & 39.0   & 36.2 &  0.945 & 0.917 \\
      &   & PC-RNN-S &   40.2 & 37.9 &  0.954 & 0.934 \\
     &   & PC-RNN-B &  \textbf{40.7}  & \textbf{39.0}  & \textbf{0.957} & \textbf{0.944}    \\
     \cmidrule(r){2-7}
     &   & CS &   31.3  & 30.9 &  0.817 &  0.792 \\
     &  & U-Net &  35.8   &  34.5 & 0.869  &  0.845     \\
     &  & VN &  36.0   & 33.8 & 0.880 &  0.840     \\
   Knee  &  & VS-Net &  36.4   & 34.5 & 0.888 &  0.854     \\
    Multi-coil &  PDFS & KIKI-Net &   36.9 & 35.5 &  0.886 & 0.862  \\
     &   & D5C5 & 37.2   & 35.5 & 0.892  &  0.865 \\
     &   & ComplexMRI & 37.3   & 35.7 &  0.893 & 0.868 \\
    &   & PC-RNN-S & 37.9   & 36.5 &  0.901 &  0.880 \\
      &   & PC-RNN-B &  \textbf{38.1}  & \textbf{37.2}  & \textbf{0.903} & \textbf{0.887}    \\
     \cmidrule(r){2-7}
     &  & CS &  31.8  &   29.9 & 0.851 & 0.808    \\
     &  & U-Net &  35.6   &  33.9  & 0.890  & 0.863 \\
     &  & VN &  36.2   & 33.4 & 0.900 &  0.857     \\
     &  & VS-Net &  36.8   & 34.4 & 0.909 &  0.875     \\
     & All  & KIKI-Net & 37.6  & 35.7 & 0.913  & 0.888  \\
     &   & D5C5 &  38.0  & 35.7 & 0.917  & 0.890  \\
     &   & ComplexMRI &  38.2  & 36.0 & 0.919  & 0.893 \\
    &   & PC-RNN-S & 39.1   & 37.2 &  0.928 &  0.907 \\
     &  & PC-RNN-B &   \textbf{39.4} & \textbf{38.1} & \textbf{0.930} & \textbf{0.916}    \\
    
    \bottomrule
  \end{tabular}
  \label{table_knee_validation_mc}
\end{table}

\begin{figure*}[!htbp]
  \centerline{
  \includegraphics[width=1.0\linewidth]{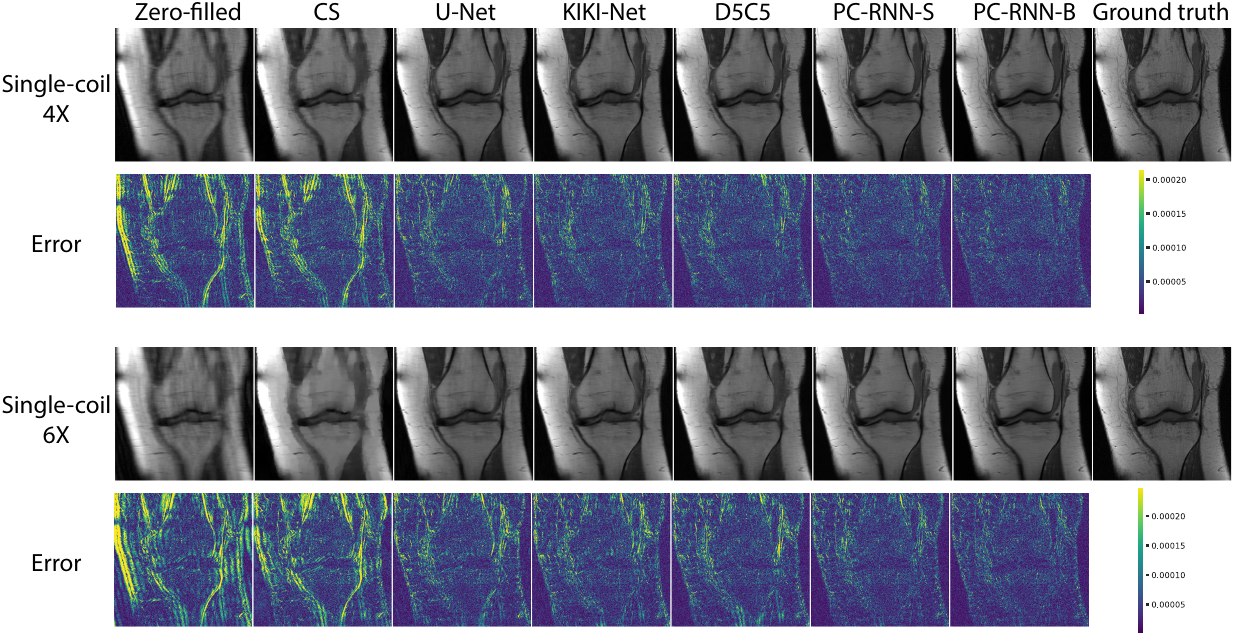}
  }
  \caption{Examples of reconstruction and error maps of knee images for the single-coil task. All errors are multiplied by 5 for better visualization.}
  \label{fig:knee_comparison_singlecoil}
\end{figure*}

\begin{figure*}[!htbp]
  \centerline{
  \includegraphics[width=1.0\linewidth]{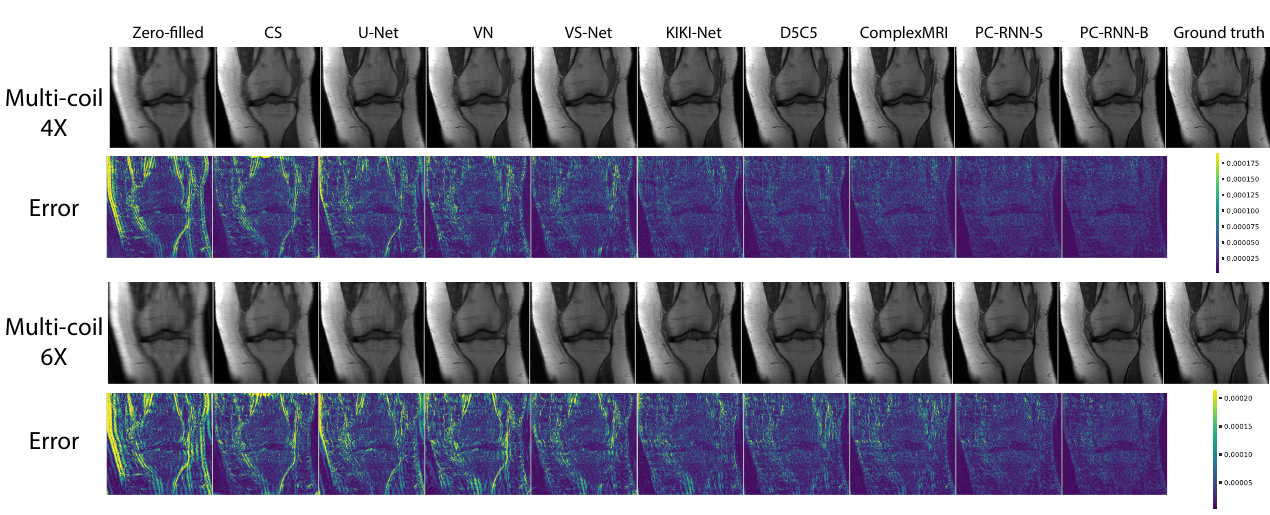}
  }
  \caption{Examples of reconstruction and error maps of knee images for the multi-coil task. All errors are multiplied by 5 for better visualization. }
  \label{fig:knee_comparison_multicoil}
\end{figure*}

We first trained and evaluated our model on the fastMRI knee dataset and compared our model with other methods. In the single-coil task (Table~\ref{table_knee_validation_sc}), both PC-RNN-B and PC-RNN-S outperform the other methods in both PSNR and SSIM at 4X acceleration. At 6X acceleration, our models achieve the best reconstruction results among all the methods  (all $p<10^{-5}$). In the multi-coil task (Table~\ref{table_knee_validation_mc}), the improvement is more significant. Comparing PC-RNN-S to the second best model ComplexMRI, PSNR is boosted by 0.9 at 4X acceleration and 1.2 at 6X acceleration. PC-RNN-B shows better performance than PC-RNN-S. SSIM also shows consistent improvement results ($p < 10^{-5}$). 

Figure~\ref{fig:knee_comparison_singlecoil} and \ref{fig:knee_comparison_multicoil} demonstrates examples of reconstructed knee images by PC-RNN-B/S and other methods. Our models recover more details, especially using multi-coil data. At 6X high acceleration, our models can reconstruct more details than the other methods.

During the 2019 fastMRI competition, we submitted the results of PC-RNN on the knee challenge dataset. All the results were first ranked by SSIM and the top four results were then evaluated by seven expert radiologists based on the following categories: contrast to noise ratio, artifacts, sharpness, diagnostic confidence, and overall image quality \cite{knoll2020advancing}. Our results ranked as one of the best in the multi-coil 4X task by radiologists' assessment. Readers who are interested in the comparison results between our model and other top models can refer to \cite{knoll2020advancing}.

\subsection{fastMRI brain dataset}

To show the generalizability of the proposed method, we also trained and evaluated PC-RNN on the fastMRI brain dataset. 

Unlike the knee dataset, where all data have 15 coils, the brain dataset contains data with various numbers of coils,  
which ranges from 2 to 28 coils. We applied coil compression to standardize the coil numbers. In order to find the best number of compressed coils, we experimented with different compressed coils and evaluated fully sampled images from coil compression with those from non-compressed coils. When the number of compressed coils is 4, the SSIM of coil-compressed images reduces to 0.931 (Table~\ref{table_coil_compression_fully_sample}). However, visual examination shows that the information loss is mainly in the background and outermost boundary regions, which is consistent with the previous report that coil compression mainly removes the background noise \cite{zhang2013coil}. 

To systematically evaluate the effect of coil compression on the reconstruction results, we trained separated PC-RNN models for the data with 4, 8 and 12 compressed coils, respectively. The PSNR and SSIM are showed in Table~\ref{table_coil_compression}. We observed that the model trained on the data with 4 compressed coils has a similar performance as the model on the data with 12 compressed coils ($p=0.62$ for PSNR and $p=0.23$ for SSIM), 
while it is slightly better than the model on 8 compressed coils ($p<0.01$ for both PSNR and SSIM). 
This experiment shows compressing the data into 4 coils has no noticeable negative impact on the reconstruction results. 

Since the brain dataset is much larger than the knee dataset and takes a much longer time to train, we decide to choose 4 compressed coils. For data with coils less than the desired number, we pad zeros on the coil dimension. Only 7 out of 5847 brain data ($0.1\%$) have coils less than 4 and need zero padding. The evaluation shows the reconstruction with 4 compressed coils achieves reasonably good results. Figure~\ref{fig:coil_compression_csm} demonstrates examples of compressed coil images and the corresponding CSM. The CSM of the four compressed coils are similar across different data regardless of the original coil settings. Therefore, the coil compression procedure standardizes the CSM of the original data and thus makes the compressed data less sensitive to the original coil configurations. Our model does not use CSM and implicitly learns the coil difference. Thus this standardization procedure also helps the model learn the coil variability when taking the compressed coils as multi-channel inputs.

\begin{table}[!tbp]
  \caption{Evaluation of coil compression on fully sampled image in brain validation dataset.}
  \centering
  \begin{tabular}{cccc}
    \toprule
    Compressed coils & PSNR  & SSIM \\
    \midrule
    4 & 73.0 &  0.931 \\
    8 & 79.6 &  0.980 \\
    12 & 89.5 & 0.996 \\
    16 & 127.5  & 0.999 \\
    \bottomrule
  \end{tabular}
  \label{table_coil_compression_fully_sample}
\end{table}

\begin{figure*}[!tbp]
  \centerline{
  \includegraphics[width=0.8\linewidth]{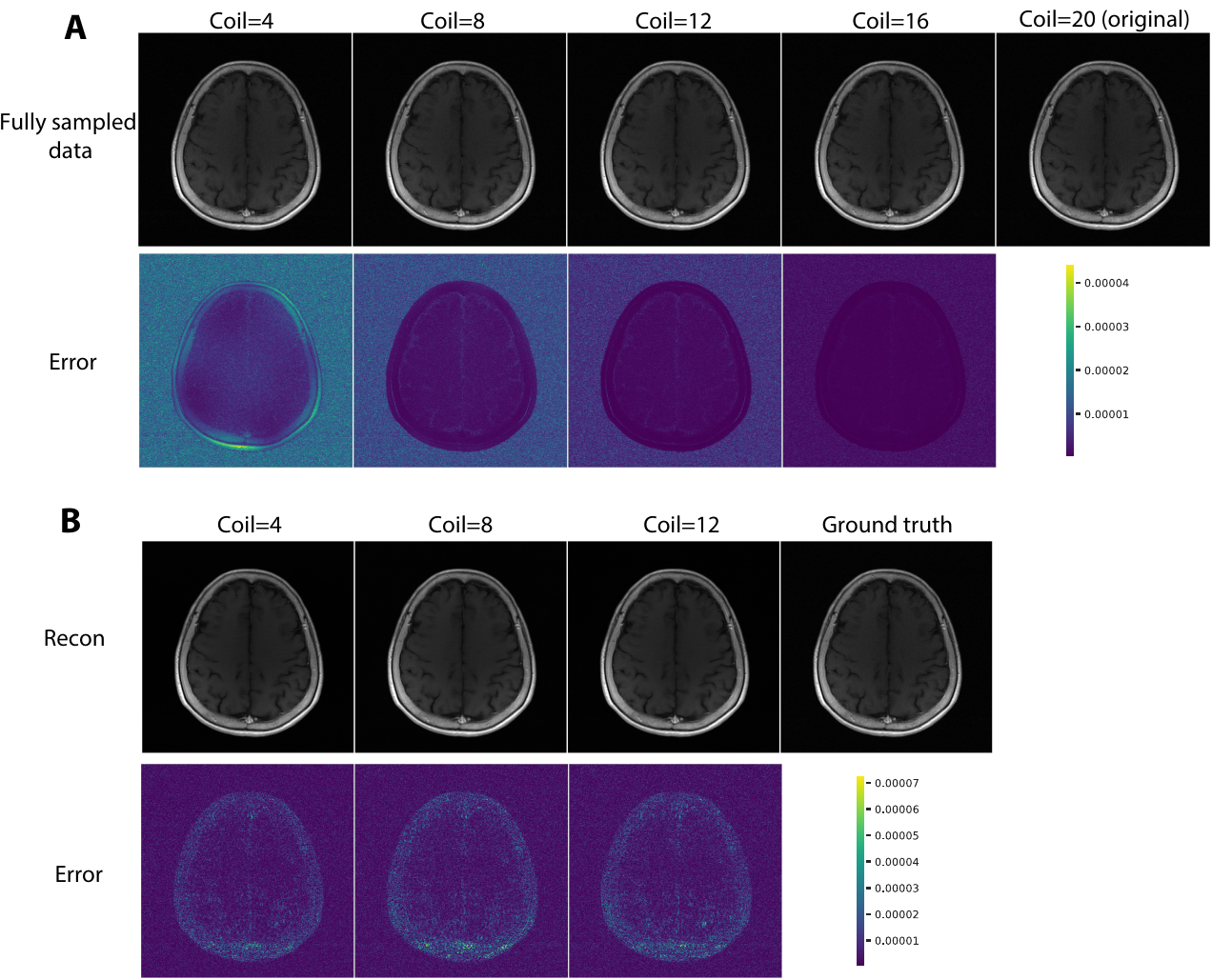}
  }
  \caption{An illustration of the coil compression with different numbers of compressed coils. (A) The fully sampled multi-coil data after coil compression. The error maps indicate the difference between the coil compressed image and the original image.  (B) The reconstruction results from three models trained with different numbers of compressed coils. The error maps show the difference between the reconstructed images and the ground truth image, which is the RSS of the original data without coil compression. All errors are at the original scale.
  }
  \label{fig:coil_compression_comparison}
\end{figure*}

\begin{table}[!tbp]
  \caption{Evaluation of PC-RNN models trained with different number of compressed coils 4X acceleration. GPU usage and running time were evaluated on a 320x320 image.}
  \centering
  \begin{tabular}{ccccc}
    \toprule
      Compressed coils    & PSNR & SSIM & GPU (train/test) & Time (train/test) \\
    \midrule
       4 & 40.8 & 0.971 & 6.8G/2.5G & 0.53s/0.22s \\
       8 & 40.4  &  0.969 & 7.8G/2.6G &  0.68s/0.24s\\
       12 & 40.7  &  0.970 & 8.8G/2.6G &  0.71s/0.25s  \\
    \bottomrule
  \end{tabular}
  \label{table_coil_compression}
\end{table}

\begin{figure*}[!htbp]
  \centerline{
  \includegraphics[width=0.8\linewidth]{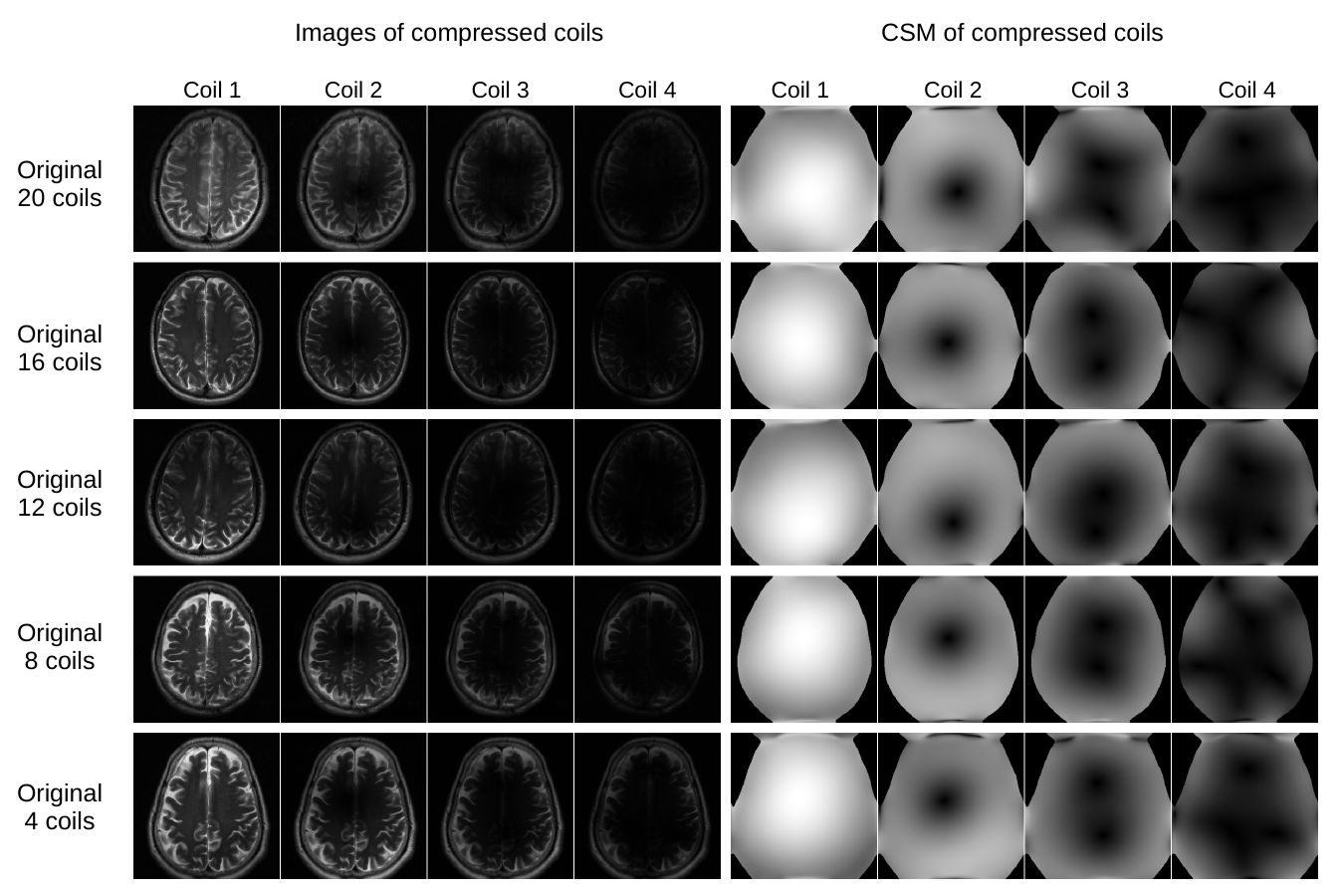}
  }
  \caption{A demonstration of the compressed coil images and the corresponding coil sensitivity maps (CSM). Data with different numbers of coils are compressed into four coils. 
  }
 \label{fig:coil_compression_csm}
\end{figure*}

\begin{table}[htbp]
  \caption{Evaluation results on the validation dataset of fastMRI brain multi-coil task.}
  \centering
  \begin{tabular}{ccccccc}
    \toprule
     &&&\multicolumn{2}{c}{PSNR}  & \multicolumn{2}{c}{SSIM}\\
    \cmidrule(r){4-5} \cmidrule(r){6-7}
    Task & Sequence  & Method &  4X     & 6X  & 4X     & 6X \\
    
    \midrule
     &    & CS & 38.6  & 34.3 & 0.918 & 0.879  \\
     &    & U-Net & 37.7 & 36.0 & 0.950 & 0.936 \\
     &    & VN & 38.7  & 35.3 & 0.952 &  0.923 \\
     &    & VS-Net & 39.9  & 36.9 & 0.960 & 0.938  \\
     &  T1& KIKI-Net & 39.4  & 37.2 & 0.960 &  0.944 \\
     &    & D5C5 &  39.3 & 36.7 & 0.957 & 0.939  \\
     &    & ComplexMRI & 39.5  & 37.2 &  0.958 & 0.942  \\
     &    & PC-RNN-S &  41.2 & 38.9 & 0.968 & 0.955  \\
     &    & PC-RNN-B & \textbf{42.2} &  \textbf{40.6} & \textbf{0.973} &  \textbf{0.965}   \\
   
   \cmidrule(r){2-7}
     &    & CS & 33.4  & 29.3 & 0.896 & 0.834   \\
     &    & U-Net & 35.4 & 33.2 & 0.941 & 0.920\\
     &    & VN & 34.8  & 31.1 & 0.930 & 0.888  \\
     &    & VS-Net &  36.4 & 32.9 & 0.946 &  0.912 \\
     &  T2& KIKI-Net & 37.0  & 34.2 & 0.953 &  0.931 \\
     &    & D5C5 &  37.2 & 33.9 & 0.953 & 0.927  \\
     &    & ComplexMRI & 36.9  & 34.1 & 0.951 & 0.928  \\
     &    & PC-RNN-S &  39.1 & 36.5 & 0.965 & 0.950  \\
     &    & PC-RNN-B & \textbf{39.9} &  \textbf{38.0} & \textbf{0.969} &  \textbf{0.960}   \\
   
   \cmidrule(r){2-7}
     &    & CS & 36.1 & 33.1 & 0.905 & 0.864  \\
     &    & U-Net & 36.4  & 	34.7 & 0.931 & 0.913 \\
     &    & VN &  37.1 & 34.0 & 0.935 & 0.900  \\
    Brain &  & VS-Net &  38.1 & 35.3 & 0.945 &  0.919 \\
    Multi-coil &  FLAIR  & KIKI-Net & 38.3  & 36.2 & 0.947 & 0.927  \\
     &    & D5C5 & 38.2  & 35.8 & 0.946 & 0.924 \\
     &    & ComplexMRI &  38.4 & 36.3 & 0.947  &  0.929 \\
     &    & PC-RNN-S &  39.8 & 37.6 & 0.956 & 0.941  \\
     &    & PC-RNN-B & \textbf{40.3} &  \textbf{38.7} & \textbf{0.960} &  \textbf{0.950}   \\
   
   \cmidrule(r){2-7}
     &   & CS & 37.8 & 33.8 & 0.918 & 0.877  \\
     &  & U-Net & 37.9 & 36.1 & 0.955 & 0.939\\
     &    & VN & 38.3  & 34.8 & 0.951 & 0.921  \\
     &    & VS-Net & 39.5  & 36.5 & 0.960 & 0.937  \\
     &  T1POST  & KIKI-Net & 39.5  & 37.1 & 0.963 & 0.947  \\
     &      &  D5C5 & 39.4 & 36.4 & 0.961 & 0.941  \\
     &    & ComplexMRI & 39.5  & 37.0 & 0.961 & 0.944  \\
     &    & PC-RNN-S &  41.4 & 38.9 & 0.972 & 0.959  \\
     &     & PC-RNN-B & \textbf{42.5} &  \textbf{40.6} & \textbf{0.977} &  \textbf{0.969}   \\
   
   \cmidrule(r){2-7}
     &      &  CS &  35.2  &  31.2 & 0.904 &  0.851   \\
     &  & U-Net &  36.3  &  34.3  & 0.944   & 0.925 \\
     &    & VN & 36.2  & 32.6 & 0.938 & 0.900  \\
     &    & VS-Net &  37.6 &  34.3 & 0.950 &  0.921 \\
     &    All   &  KIKI-Net &   37.9 & 35.3  &  0.956 &  0.935  \\
     &      &  D5C5 &  38.0  &  34.9 & 0.955 &  0.931  \\
     &    & ComplexMRI & 37.9  & 35.2 & 0.953 &  0.933 \\
     &    & PC-RNN-S &  39.9 & 37.4 & 0.966 & 0.952  \\
     &  & PC-RNN-B & \textbf{40.8} &  \textbf{38.9} & \textbf{0.971} &  \textbf{0.962}   \\
    \bottomrule
  \end{tabular}
  \label{table_brain_validation}
\end{table}

\begin{figure*}[htbp]
  \centerline{
  \includegraphics[width=1.0\linewidth]{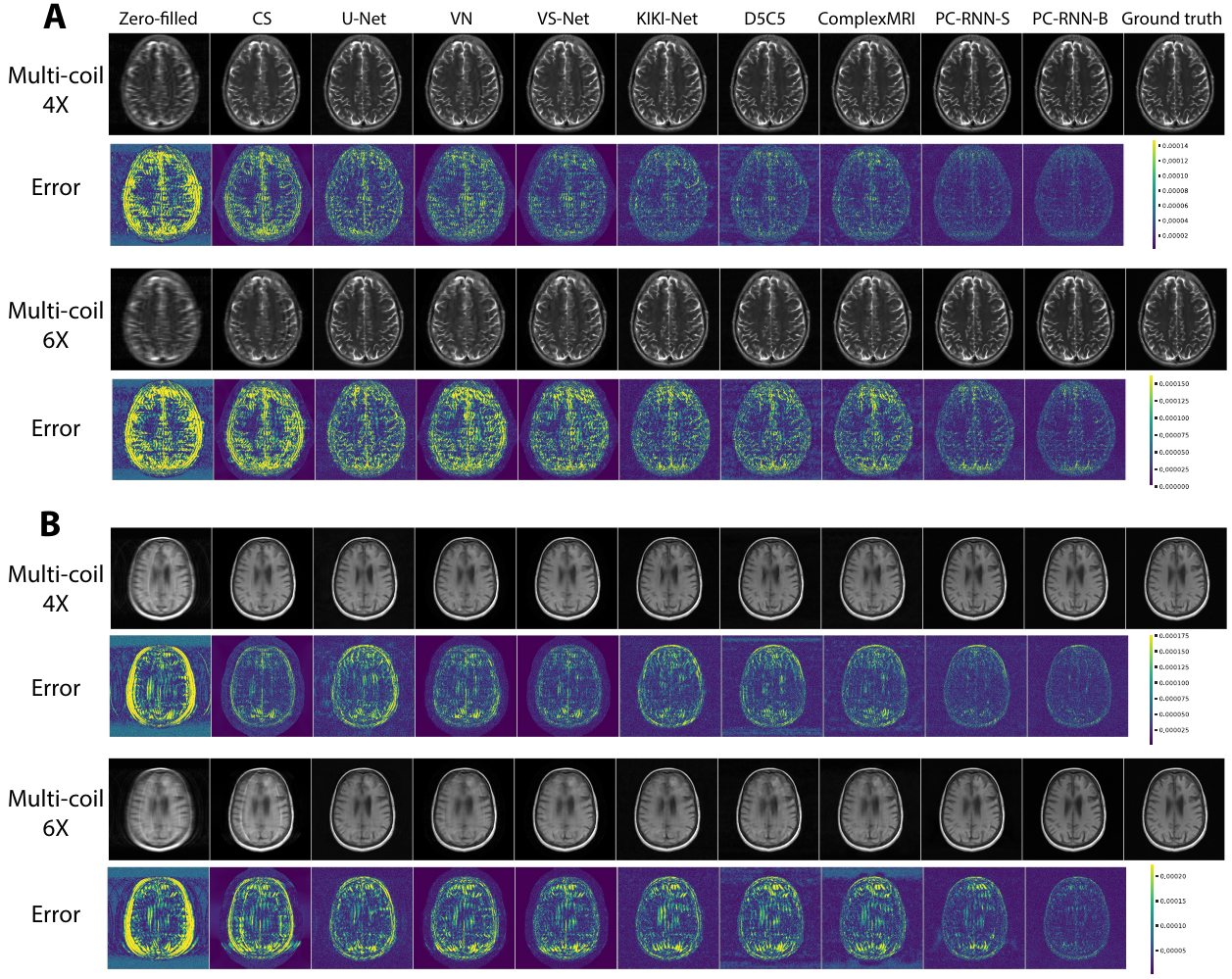}
  }
  \caption{Examples of reconstructed brain images and error maps for (A) T2-weighted and (B) T1-weighted MRI. All errors are multiplied by 5 for better visualization. 
  }
  \label{fig:brain_comparison}
\end{figure*}

Table~\ref{table_brain_validation} shows the evaluation of reconstruction results on the brain validation dataset. Similarly, both PC-RNN-B and PC-RNN-S show improvements over other methods (all $p < 10^{-5}$). The reconstructed images and error maps in Figure~\ref{fig:brain_comparison} demonstrates the improved image quality by PC-RNN-B and and PC-RNN-S, especially the fine details. The large version PC-RNN-B shows a better performance than the small version PC-RNN-S ($p < 10^{-5}$). These results show that our method outperforms other methods on different anatomies and MRI contrasts.

\begin{table}[h]
  \caption{Evaluation of outputs from each ConvRNN module of PC-RNN on fastMRI knee and brain validation datasets.}
  \centering
  \begin{tabular}{cccccc}
    \toprule
     &&\multicolumn{2}{c}{PSNR}  & \multicolumn{2}{c}{SSIM}\\
    \cmidrule(r){3-4} \cmidrule(r){5-6}
    Task &  Output    & 4X     & 6X  & 4X     & 6X \\
    \midrule
      &  ConvRNN$_{1}$-4x & 30.5 & 28.6 & 0.795 &  0.742  \\
    Knee  &  ConvRNN$_{2}$-2x & 36.2 & 34.6 & 0.897 &  0.876  \\
    Multi-coil  &  ConvRNN$_{3}$-1x & 26.7 & 22.3 & 0.738 &  0.619   \\
      &  PC-RNN &  \textbf{39.4} & \textbf{38.1} & \textbf{0.930} & \textbf{0.916}   \\
    \midrule
      &  ConvRNN$_{1}$-4x &  34.5 & 33.3 & 0.919 & 0.903  \\
    Brain  &  ConvRNN$_{2}$-2x &  34.7 & 33.4 & 0.913 &  0.897  \\
    Multi-coil  &  ConvRNN$_{3}$-1x &  32.2 & 28.9  & 0.870 & 0.784  \\
    &  PC-RNN &  \textbf{40.8} &  \textbf{38.9} & \textbf{0.971} &  \textbf{0.962}   \\
    \bottomrule
    
  \end{tabular}
  \label{table_multiscale}
\end{table}

\begin{table}[ht]
  \caption{Ablation study of PC-RNN on fastMRI knee and brain validation dataset at 4X acceleration. The arrows indicate the order of ConvRNN modules.}
  \centering
  \begin{tabular}{cccc}
    \toprule
    Task &  Ablated PC-RNN    & PSNR & SSIM \\
    \midrule
      &  4x & 38.4 & 0.920  \\
      &  2x & 39.0 & 0.927  \\
      &  1x & 38.6 & 0.923  \\
      \cmidrule(r){2-4}
    Knee   &  4x$\rightarrow$2x & 39.2 & 0.928  \\
    Multi-coil  &  4x$\rightarrow$1x & 38.9 & 0.925  \\
      &  2x$\rightarrow$1x & 39.1 & 0.927  \\
      \cmidrule(r){2-4}
      &  4x, 2x, 1x &  39.1  &  0.928 \\
      &  4x$\rightarrow$4x$\rightarrow$4x &  38.5 & 0.921 \\
      &  1x$\rightarrow$2x$\rightarrow$4x &  38.8 & 0.925 \\
      &  4x$\rightarrow$2x$\rightarrow$1x &  \textbf{39.4} & \textbf{0.930} \\
    \midrule
      &  4x & 39.5 & 0.965  \\
      &  2x & 39.6 & 0.965  \\
      &  1x & 38.6 & 0.960  \\
      \cmidrule(r){2-4}
    Brain   &  4x$\rightarrow$2x & 39.9 & 0.967  \\
    Multi-coil  &  4x$\rightarrow$1x & 39.7 & 0.966  \\
      &  2x$\rightarrow$1x & 39.8 & 0.966  \\
      \cmidrule(r){2-4}
      &  4x, 2x, 1x &  40.0 & 0.967 \\
      &  4x$\rightarrow$4x$\rightarrow$4x &  39.7 & 0.966 \\
      &  1x$\rightarrow$2x$\rightarrow$4x &  39.9 & 0.967 \\
      &  4x$\rightarrow$2x$\rightarrow$1x &  \textbf{40.8} & \textbf{0.971} \\
    \bottomrule
  \end{tabular}
  \label{table_ablation}
\end{table}

\begin{figure*}[!htbp]
  \centerline{\includegraphics[width=1.6\columnwidth]{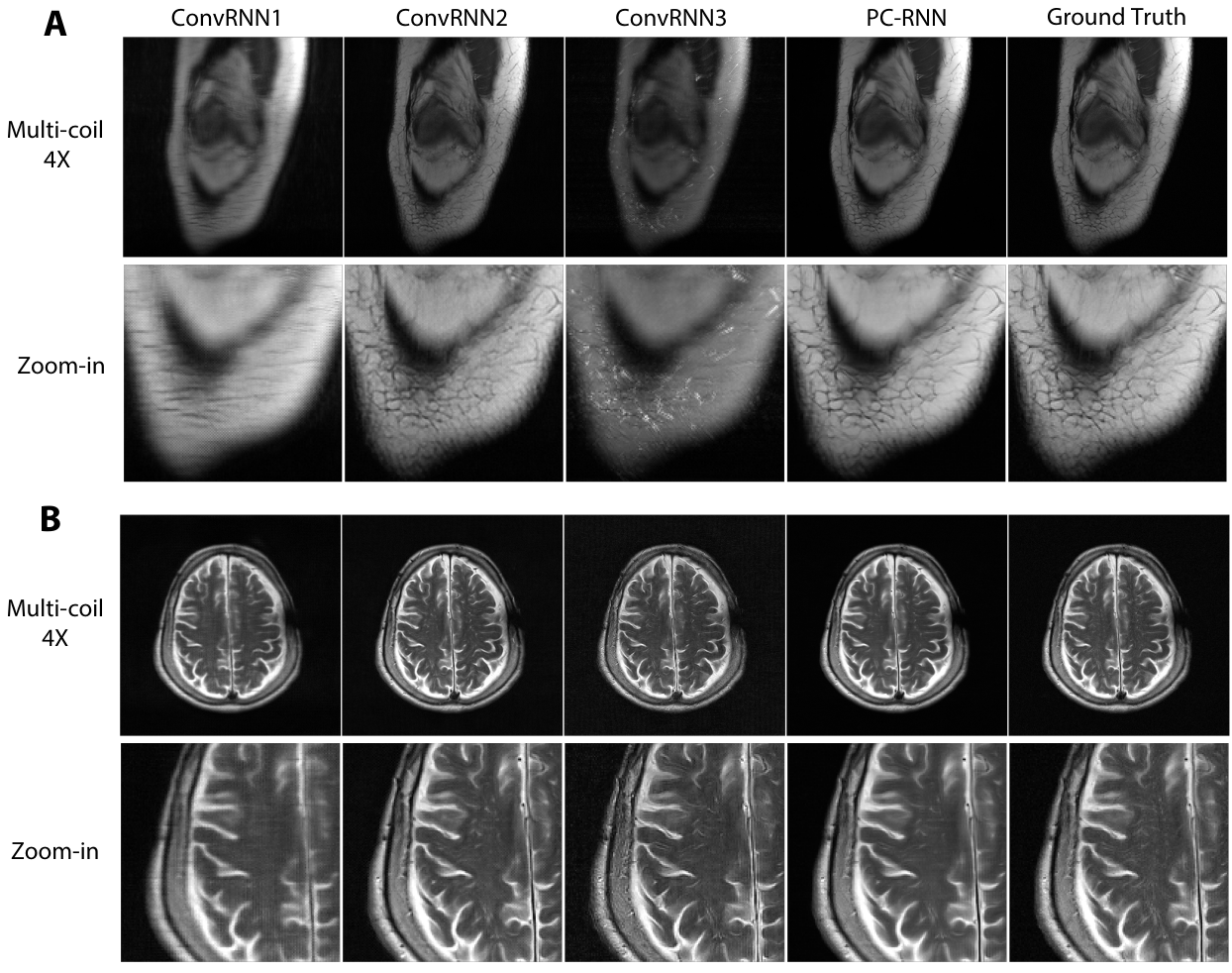}}
  \caption{Intermediate and final outputs of PC-RNN for (A) knee images and (B) brain images. Outputs from ConvRNN$_1$, ConvRNN$_2$ and ConvRNN$_3$ show that PC-RNN learns to reconstruct images from coarse to fine scales. The details in the output of ConvRNN$_3$ are enhanced. The final reconstruction that combines the multi-scale reconstructions depicts a clear and sharp image, which is similar to the ground truth. }
  \label{fig:pc_rnn}
\end{figure*}

\subsection{Multi-scale learning in PC-RNN}
To demonstrate the effectiveness of PC-RNN in learning multi-scale features, we show the intermediate outputs from each ConvRNN as well as the final reconstructed image and compared them with the ground truth in Figure~\ref{fig:pc_rnn}. From ConvRNN$_1$-4x to ConvRNN$_3$-1x, additional fine structures are recovered in the reconstructed images. For both knee and brain results, ConvRNN2 outputs are sharper than the ConvRNN1 outputs, indicating more high frequency information is learned by ConvRNN2. The ConvRNN3 outputs depict more enhanced details when comparing to the outputs of ConvRNN1 and ConvRNN2. This is in agreement with the network design that ConvRNN3 should be able to explicitly learn the high frequency details. Interestingly, ConvRNN3 seems to learn different high frequency features for knee and brain data, respectively. This is probably because the two datasets have different image contrasts as well as resolutions and therefore the model learns to utilize different features for knee and brain datasets. The final CNN in PC-RNN combines all three reconstructed images and outputs a clear and sharp MRI image, which is more similar to the ground truth image. 

Table~\ref{table_multiscale} compares the intermediate outputs against the ground truth on the knee and brain datasets. It is interesting to observe that each ConvRNN generates a relatively low-quality image but the final reconstructed image based on those intermediate images achieves much better results. This is because the final CNN module aggregates all the features from the three ConvRNN modules and provides guidance for ConvRNN modules to extract features at different scales. Therefore, each ConvRNN module may focus on learning complementary features such that the final CNN module generates the best result, which is also indicated by the results in Figure 7.

We performed ablation studies to demonstrate the effect of multi-scale reconstruction in our model (Table~\ref{table_ablation}). The ablated models are all trained from scratch on the knee and brain datasets at 4X acceleration. First, we removed ConvRNN modules at certain scales. The PC-RNN model with all three scales (4x$\rightarrow$2x$\rightarrow$1x) performs better than the ablated models with a single scale (1x, 2x or 4x) and models with two scales (2x$\rightarrow$1x, 4x$\rightarrow$1x or 4x$\rightarrow$2x) (all $p<10^{-5}$). we also trained a single-scale model with three ConvRNN1 modules (4x$\rightarrow$4x$\rightarrow$4x) and the performance is worse than the original multi-scale PC-RNN model. Note that ablated PC-RNN models with a single scale are just ConvRNN models without multiple scales. This result illustrates the benefits of multi-scale learning. To further demonstrate the advantage of the sequential coarse-to-fine reconstruction, we also trained a model with a reversed multi-scale architecture such that images are reconstructed in a fine-to-coarse way (1x$\rightarrow$2x$\rightarrow$4x). Additionally, we trained another ablated model by removing all the connections between ConvRNN modules (4x,2x,1x), where each ConvRNN takes the undersampled image as the input and extracts features at different scales but in a parallel fashion. Both models perform worse than the original PC-RNN model (all $p<10^{-5}$), which indicates that the sequential coarse-to-fine reconstruction is beneficial.

\section{Discussion}
One major difference between our method and other published methods is that our model learns features in a coarse-to-fine manner using a pyramid architecture. The proposed PC-RNN model includes three ConvRNN modules and each has the recurrent encoder-decoder structure with different downsampling rates, which project the data into lower dimensions at different rates. Different from other methods, our multi-scale model learns to reconstruct MRI images based on low, middle and high frequency information in a sequential pyramid manner. The PC-RNN architecture helps the model to learn the high frequency information based on the low frequency information, which leads to better recovery of fine details than other methods. In the current model, we only implemented three scales (1x, 2x, and 4x). The reconstruction can be further improved by including more scales in the proposed framework. Another major difference with other CNN based method is that our method is an RNN based model. We discuss the detailed difference between some previous methods utilizing high frequency learning and our method as follows.

Zhang \textit{et al.} \cite{zhang2020compressed} proposed a cascaded CNN model together with frequency loss. First, our proposed PC-RNN is an RNN based model that is designed to learn optimization. Second, their proposed method employs a  frequency loss, which is a combination of high frequency loss and low frequency loss. Both losses are added to every intermediate output of their model and thus intermediate reconstructions are not in a sequential coarse-to-fine manner. Our model forms a pyramid structure, where the coarse reconstruction results are fed into the next module as inputs for a finer reconstruction and the reconstructions at different scales are then aggregated to generate the final result. The model reconstructs images in a sequential coarse-to-fine way, which has been demonstrated in Figure~\ref{fig:pc_rnn}.  

Nakarmi \textit{et al.} \cite{nakarmi2020multi} proposed a CNN (ResNet) based unrolled deep learning model to reconstruct images at two scales. They first processed the input image with a low pass filter and a high pass filter to generate the low resolution image (LR) and high resolution (HR) image, respectively. The LR and HR images are then fed into two ResNet models for reconstruction and the final results are combined to generate the final reconstruction. The LR and HR images are reconstructed in a parallel fashion and thus the HR reconstruction does not utilize the LR results. The proposed PC-RNN model does not need the manual processing of the input images with low/high pass filters. Instead, the three ConvRNN modules explicitly learn how to extract important features at different scales for reconstruction. Also, the previous coarse scale reconstruction results are utilized by the following fine scale reconstruction modules in a pyramid fashion. The ablation studies in Table~\ref{table_ablation} demonstrates the benefits of reconstruction in such a sequential coarse-to-fine manner compared to the parallel fashion.

Liang \textit{et al.} \cite{liang2020laplacian} developed a cascaded multi-scale network for MRI image reconstruction.  The CNN based network includes two branches for Laplacian decomposition and shuffle downsampling, respectively. The shuffle downsampling branch first compressed the data into a smaller scale, which is then fed into three sub-branches to extract features at different scales separately. The multi-scale features are then fused with the Laplacian error maps and Gaussian map generated from the Laplacian decomposition branch. By contrast, the proposed PC-RNN is an RNN based model. The multi-scale features in PC-RNN are learned in a sequential coarse-to-fine manner, which can be viewed as a multi-scale searching strategy that often leads to faster convergence and better solutions than directly optimizing in the original large space \cite{mjolsness1991multiscale}. 

Dar \textit{et al.} \cite{dar2020prior} proposed a GAN based model to learn the low frequency and high frequency priors, which are then utilized by the generator to synthesize reconstructed images. Since GAN is a generative model, such methods may introduce undesired artifacts \cite{qin2020deep,tan2021systematic,abdal2019image2stylegan,demir2018patch,zhu2020gan,lucas2019efficient,chen2021hierarchical}, which is unfavorable in the clinical setting. Furthermore, like the other methods, the low frequency and high frequency priors in \cite{dar2020prior} are learned separately. In the proposed method, the high frequency features are learned based on the previously learned low frequency features and the multi-scale information is combined to generate the final result. Such pyramid network architecture has been shown to have many benefits in several computer vision applications involving small objects \cite{adelson1984pyramid,lin2017feature,zhao2017pyramid,fu2019lightweight,he2015spatial}. 

The residual learning has been employed in many MRI reconstruction methods, such as D5C5\cite{schlemper2017deep}, KIKI-Net, VS-Net and PC-RNN (the lower right part in Figure 1 shows the ResBlock with two residual connections). The residual learning enforces the network to learn the high frequency information and thus helps to reconstruct details. We utilize this idea inside the ResBlock of PC-RNN model. As we explained above, different from those previous works using residual learning, the proposed method also learns to reconstruct images at different scales in a pyramid fashion. 

The benefits of directly taking coil compression data as multi-channel inputs than utilizing CSM are three folds. First, the coil compression removes the background noise \cite{zhang2013coil} and thus can potentially help the training converge faster. Second, it standardizes the coils since after SVD, the compressed coils are the principal components in the coil dimension and the compressed coils are standardized by their corresponding eigenvalues (Figure~\ref{fig:coil_compression_csm}). Thus, the original coil configuration is less relevant and the relationship of compressed coils is easier to model. Third, it speeds up the reconstruction and dramatically reduces data storage. By comparing 12 and 4 compressed coils with a batch size of one, the training GPU memory and time decrease from 8.8G and 0.71 sec/sample to 6.8G and 0.53 sec/sample, respectively. In the CSM approach, since data with different coil dimensions can not be loaded into the same batch, the batch size can only be one for training and testing, which reduces the overall training and testing speed. Furthermore, the running time reported in Table~\ref{table_coil_compression} does not include the CSM computation or coil compression, so the overall difference will be more significant if considering data preprocessing. For instance, the coil compression approach avoids the CSM calculation and is about four times faster than the CSM calculation in our case (0.53s vs. 2.08s, benchmarked on data with 15 coils and image size 320x320). Also, the  Fourier transform  is  about  five times  faster  on  coil  compressed  data  than  original  data  used  in  CSM  approach (0.02s vs. 0.10s, benchmarked on the same data).  Note  that  in  the  data  preprocessing  as well as  data consistency steps, there are often multiple FFT/iFFT operations. Therefore the time difference due to multiple FFT/iFFT will be more dramatic between the CSM approach and the coil compression approach. Also, the above two experiments were performed on one 2D image and the fastMRI brain dataset has 70,748 2D slices for training and 21,842 for testing, hence the difference in running time between these two approaches will be more significant on the whole dataset. Also, the size of the brain dataset is reduced from 1.9T to 0.4T after coil compression. Considering the CSM approach is currently the most common method for parallel imaging, our method provides an alternative and efficient way to handle multi-coil data using deep learning models. 

In this study, we performed coil compression as a data preprocessing step. Alternatively, an SVD layer can be implemented as the first layer of the PC-RNN model to perform coil compression inside the model. The SVD layer just performs the matrix decomposing of the input and therefore does not introduce any trainable parameters. Then the model itself can take data with any number of coils. Furthermore, the coil compression procedure can be considered as a pre-learned convolutional layer with a 1$\times$1 kernel. Each column of the compression matrix is a filter and there are a total $n_{vc}$ filters, which convert the input $n_c$-channel data to $n_{vc}$-channel features. Therefore, the coil compression procedure can also be integrated into the deep learning model as a learnable layer.

One possible downside of coil compression is that it results in information loss, which may include both useful signal and undesired noise. At least on the fastMRI brain data, the impact of signal loss from coil compression is not significant, while removing the noise can potentially help the model converge faster. On the other hand, the choice of the desired number of compressed coils also needs to be optimized for different data and applications. In this study, we chose the 4 compressed coils for the fastMRI brain dataset due to the consideration of computational cost and the reconstruction performance and it worked reasonably well.

\section{Conclusion}
In this paper, we proposed a multi-scale pyramid convolutional RNN model for MRI image reconstruction. Although we demonstrated our method on MRI image reconstruction, it can be extended to other imaging modalities such as CT image reconstruction. The basic idea of learning multi-scale features for inverse problems as well as decomposing the original search space into several smaller ones can be generalized to other inverse problems such as image super-resolution or inpainting, which will be among our future work.

\bibliographystyle{IEEEtran}
\bibliography{ref}

\end{document}